%% file: exo_chap.tex
\documentclass{cplslarge}

\usepackage[authoryear]{natbib}
\usepackage{color}
\usepackage{makeidx}
\makeindex

\include{chocros}

\DeclareUnicodeCharacter{2212}{-}
\DeclareUnicodeCharacter{2009}{FIXME}

\begin{document}

\setcounter{chapter}{4}


\author[Cho, Thrastarson, Koskinen, Read, Tobias, Moon \&\ Skinner]{J.Y-K. Cho,
  H.Th. Thrastarson, T.T. Koskinen, P.L. Read, S.M. Tobias, W. Moon, and
  J.W. Skinner}

\chapter{Exoplanets and the Sun}

\section{INTRODUCTION}\label{intro}

In this chapter, we review the recent progress in understanding the jet
structures on exoplanets and for the Sun.  For the former, the primary focus is
on hot-Jupiters, given that many more observations are available for them
presently \citep[see, e.g.,][]{Cho08,Showetal11,HengShow15}.  Here we make no
attempt to be comprehensive, emphasizing only the more robust aspects of
observation and numerical modeling---and, in particular, those that relate more
directly to jets, the topic of this book.  Even so, we apologize at the outset
for not including all of the voluminous studies that may be relevant to jets of
the exoplanets and the Sun.  In addition, the views expressed primarily reflect
the authors' own understanding and biases, and may not be entirely in line or
agree with those of the community at large presently.  Hence, the readers are
encouraged to use the present chapter as a starting point for their own
exploration and analyses.

Before embarking on our discussion, the first thing we note is that, because we
are still in the early stages of observation and theoretical modeling of
exoplanets, not much can actually be said about the presence and strength (let
alone the morphology) of the jets on exoplanets.  Accordingly, as discussed
below, there is not much agreement at this point---in observations or in
theory.  The picture is very different for the Sun, for which it can be argued
the jet structure is very well known.  In fact, the jet structure of the Sun is
arguably one of the best known jet structures of all the planets and stars.  As
we shall see, the situation is due to the fact that Sun's disk is resolved and
its interior can be probed with helioseismology
\citep[e.g.,][]{GougToom91,Chri-Dals02,MiesToom09}.

The outline of this chapter is as follows.  In Section~\ref{exo}, we begin with
a summary of recent observations and simulations of  temperature distribution
(in particular, `hot spot' location) and variability, as they relate to the
possible presence of a high-speed equatorial jet on exoplanets
(Sections~\ref{exo_obs} and \ref{exo_sim}, respectively).  Then, a brief
discussion of several critical issues pertaining to numerical modeling of jets
on exoplanets follows (Section~\ref{exo_iss}).  In Section~\ref{sun}, this basic
structure is repeated for the Sun, with brief summaries of the observations and
three-dimensional (3D) global numerical simulations (Sections~\ref{sun_obs} and
\ref{sun_sim}, respectively).  This in turn is followed by a discussion of some
crucial issues for modeling and understanding the jets on and in the Sun
(Section~\ref{sun_iss}).  In Section~\ref{discuss}, we conclude with a brief
summary of the chapter and present a broad outlook on the subject of jets for
both exoplanets and the Sun.  The discussion in the present chapter can be
profitably followed concurrently with the closely-related materials in Chapters
2, 4, and 10 of the book.

\section{EXOPLANETS}\label{exo}

\subsection{Current Observations}\label{exo_obs}

At present, observations of ``surface'' (i.e., atmospheric) features on
exoplanets are extremely challenging.  This is because the planetary disk is not
resolved, unlike the planets in the Solar System.  Accordingly, interpretation
of acquired observations are highly controversial.  For example, there are
numerous cases of observations of the same planet at different epochs which
disagree significantly (sometimes by the same group!).  While this may be
indicating an intrinsic variability of the planet, the variations are primarily
attributed by the observers to variability in the instrument and/or data
handling.

The significance of variability for exoplanets is in its putative connection
with a high-speed prograde (eastward) equatorial jet, often produced in
numerical simulation studies
\citep[e.g.,][]{Choetal08,Showetal09,RausMeno10,Hengetal11,Maynetal14,
  Tsaietal14,Choetal15}.  In many of these studies, the jet is supersonic---even
hypersonic, and it exhibits very little zonal (longitudinal) and temporal
variabilities.  Several theoretical studies, however, have emphasized the
possibility of wave-induced variability
\citep[e.g.,][]{Choetal03,Choetal08,WatCho10,Choetal15} as well as
vortex-induced variability \citep[e.g.,][]{Choetal03,ThraCho10}---stressing the
coupling between the variability and the jet, rather than the variability simply
being dictated by the jet.  Only recently one case of observed variability has
not been disputed, thus far.  Currently, there is no direct observational
evidence of jets on an exoplanet.

The situation with {\it HD189733b}, one of the best observed exoplanets so far,
presents a good illustration of the disagreements.  \citet{Griletal08} report
discrepancies in spectral emission features between two
epochs.  \citet{Charetal08} find spectral features in secondary eclipse that are
inconsistent with measurements obtained earlier by \citet{Griletal07}.  At the
same time, \citet{Charetal08} find a different eclipse depth from that reported
by \citet{Knutetal07}.  On the other hand, \citet{Singetal09} obtain
measurements at two wavelengths during a transit and report that they do not
observe the water signature previously reported by \citet{Swaietal08}.  In
\citet{Agoletal10}, six pairs of transits and eclipses of {\it HD189733b} are
studied, concluding with ``1-$\sigma$'' (68\% confident) constraints on the
variability consistent with a low variability on the dayside and a high
variability on the nightside.  Further, observational results of
\citet{Knutetal12} differ from those of \citet{Charetal08}, with the former work
expressing that the values of the latter work require revision.

The same state of affairs exists for {\it HD209458b}, another of the best
observed exoplanets.  \citet{Crosetal12} report a mean transit depth consistent
with previous measurements, showing no evidence of variability in transit depth
at the 3\% level, but they report a mean eclipse depth somewhat higher than that
previously reported for this planet.  In a more recent observation by
\citet{Zelletal14}, they report inconsistency with an earlier study by
\citet{Knutetal08}, but suggest that the inconsistency is due to variation in
the instrument usage between the two studies.  In addition, \citet{Zelletal14}
revise their previous 4.5\,$\mu$m measurement of {\it HD209458b}'s secondary
eclipse (i.e., when the planet is behind its host star from the point of view of
the observer) emission downward by approximately 35\%, a very large
amount.  They also report that the dayside brightness temperature is
$1499 \pm 15$\,K and the nightside emission temperature is $972 \pm 44$\,K for
this planet, a difference of nearly 500\,K that is broadly consistent with the
setup used in many simulations.

In contrast, \citet{Crosetal10} observe a non-transiting giant exoplanet, {\it
  ups\,And\,b}, and find the phase curve amplitudes at two different epochs to
be consistent to 1.7$\sigma$ certainty.  This is an example of non-variability
(in time).  An example of variability in time that has {\em not} experienced
dispute so far has recently been put forward for the hot-Jupiter {\it HAT-P-7b}
by \citet{Armsetal16}.  For this planet, \citet{Armsetal16} report variations in
the phase curve peak offset with the peak brightness repeatedly shifting from
east side of the planet's substellar point to the west side.  The existence of
such a behaviour has been previously shown by \citet{Choetal08} in their
simulation study (see Figure~13 therein).

Note that {\it HAT-P-7b} is an extremely hot planet with radius 1.4\,$R_J$,
where $R_J$ is the radius of Jupiter ($7.1 \times 10^7$\,m), and with a dayside
brightness temperature of $\sim$2,860\,K and an equilibrium temperature of
$\sim$2,200\,K.  It transits its host star with a period of $\sim$2.2\,days and
has been continuously observed for four years by the Kepler telescope at optical
wavelengths.  It has also been intensively observed at infrared wavelengths with
the Spitzer telescope.  Hence, in principle, it is one of the best objects to
focus on for modeling work.  However, observational studies of this planet have
suffered from the usual disagreements with the measured amplitudes, as well as
with the albedos and temperatures derived from them.  The disagreements, here
again, have been attributed to the differing wavelengths, datasets, and analysis
methods used.  Nevertheless, observations appear to agree that the variability
occurs on a timescale of tens to hundreds of days for this planet.

Interestingly, several full-phase observations of the exoplanets, {\it
  HD189733b} \citep{Knutetal07}, {\it HD209458b} \citep{Zelletal14}, and {\it
  WASP-43b} \citep{Stevetal14}, report that a ``hot spot''---thermally brightest
region on the disk---to be eastward of the substellar point.  Some circulation
model studies \citep[e.g.,][]{Showetal09, Kataetal15} result in a fast
equatorial eastward jet and the hottest region eastward of the substellar point,
which is used to claim a rough ``agreement'' (in the degree of offset).  In
turn, this has been interpreted in some observational studies as ``observational
confirmation'' that hot-Jupiters have fast (i.e., at least several kilometers
per second) zonal winds and a strong equatorial jet---even though there is
limited or no information about the latitudinal position of the brightest region
and even though other dynamic mechanisms could potentially result in the
brightest region being eastward of the substellar point.

For example, \citet{Zelletal14} report that the hot spot is shifted eastward of
the substellar point by $40.9 \pm 6$ degrees, asserting that this is in ``rough
agreement''~[{\it sic}] with circulation models predicting equatorial
superrotation; this is despite the fact that many of them have shifted hot spots
concomitant with a supersonic jet resulting from solving the primitive equations
with free-slip boundary conditions \citep[e.g.,][]{Showetal09}, which is not
physically valid \citep[e.g.,][]{Choetal15}.  Both the simple distribution and
location of the putative hot spot, as well as the robustness of obtained
numerical simulation results, have been questioned in several studies
\citep[see, e.g.,][]{Choetal08,ThraCho10,Choetal15}.  More recently, Kepler
observations show that the light-curves of some hot-Jupiters are
``asymmetric''---i.e., for the hottest planets, the light-curve peaks {\em
  before} the secondary eclipse, whereas for planets cooler than $\sim$1900\,K,
it peaks {\em after} the secondary eclipse.

For objects other than hot-Jupiters, the evidence of surface variability is
thought to be somewhat more clear and numerous.  For example, \citet{Demoetal16}
report a ``4$\sigma$-detection'' of variability in the dayside thermal emission
from the transiting super-Earth {\it 55\,Cancri\,e}.  Here the signal varies by
a factor of 3.7---a very large variation.  There is also ample evidence of
variability in brown dwarf atmosphere, based on observations in the current
literature; see, for example, \citet{Artietal09}, \citet{Radietal12}, and Apai
et al. (submitted).  It is common, according to the survey of
\citet{Buenetal14}.  However, no direct observational evidence of jets on
super-Earths or brown dwarfs have been obtained so far as well.

As already noted, although the disagreements have been attributed to variability
in the instrument and data handling, \citet{McCuetal14} argue that  the slope in
the short wave spectral region observed could be due to starspots.  Note that
the unocculted spot coverage is unknown.  However, starspots do not account for
the lack of the alkali metal wings in the data or for the fact that the slope
appears at a higher effective altitude than the  1.4\,$\mu$m water feature,
bringing into question the role of the spots.

\subsection{Review of Recent Simulations}\label{exo_sim}

In general, observations currently rely heavily on simulations for interpreting
the obtained data.  Unfortunately, as in observations, current simulation
studies are not free from discrepancies and disputes---although arguably less
than in observations.  Thus far, simulation results roughly fall into two
general categories: {\it i}) those which exhibit supersonic equatorial jets,
with hot spots shifted eastward of substellar point by some finite angular
distance, and no variability \citep[e.g.,][]{Showetal09}; and, {\it ii}) those
which exhibit subsonic equatorial jets in either direction with complex
distribution of, not-necessarily steady, hot (and cold) regions, and often with
pronounced variability and noticeable dependance on initial condition
\citep[e.g.,][]{Choetal08,ThraCho10}.  Note that the two categories are not
``mutually exclusive''.  For example, an eastward-shifted hot spot can also be
present with a subsonic equatorial jet.

\begin{figure} 
  \figurebox{20pc}{}{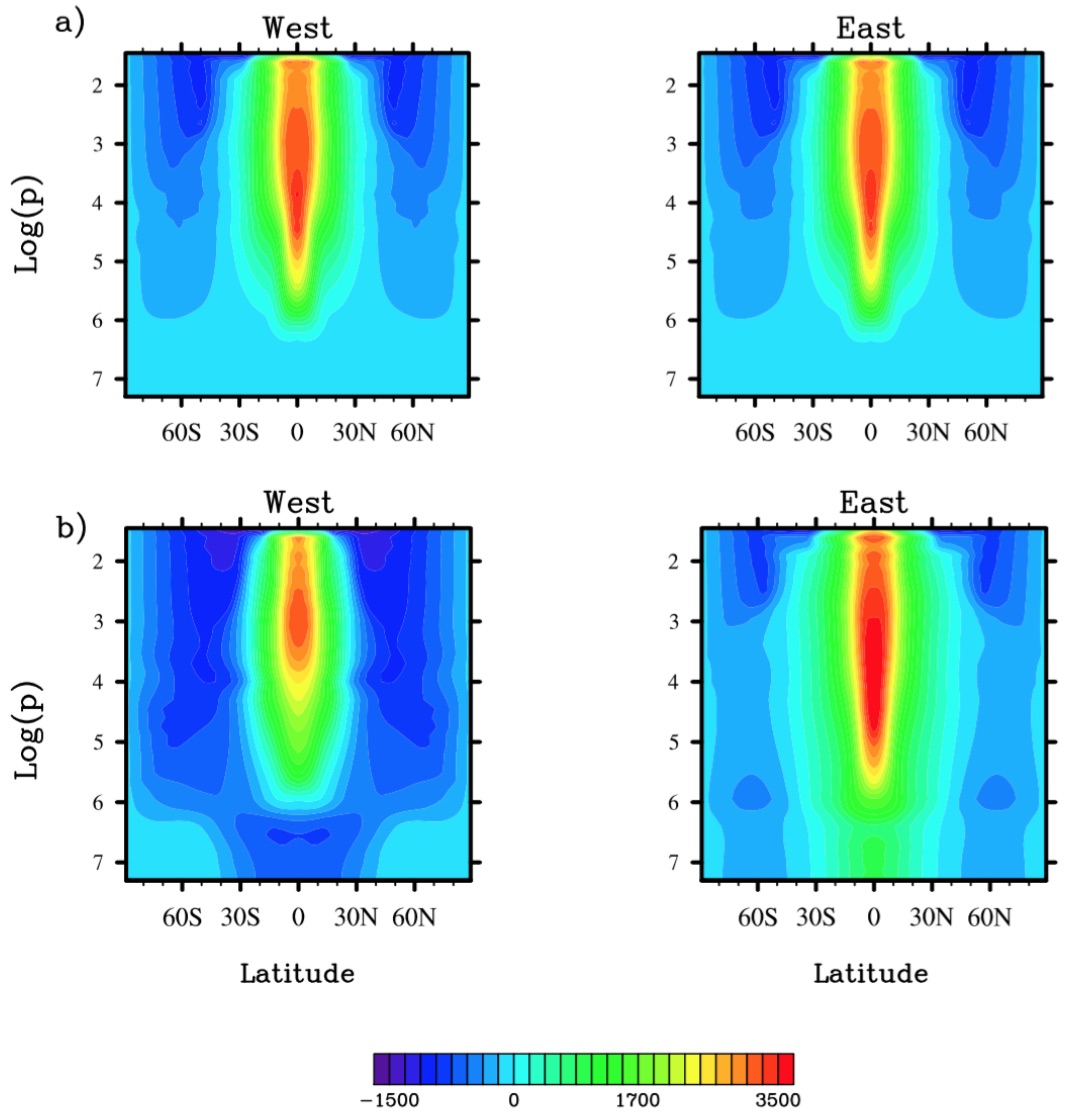}
  \caption{Time-averaged, zonal-mean, zonal velocity $\bar{u}(\vartheta, p)$ in
    m\,s$^{-1}$, where $\vartheta$ is the latitude (in degrees) and $p$ is the
    pressure (in Pa).  Simulations are performed with the PEBOB general
    circulation model, initialized with $\pm 1000$\,m\,s$^{-1}$ (East/West)
    equatorial jets and with vertical layers equally spaced in $\log(p)$.  In
    a), the bottom momentum drag timescale is 3.3$\tau$, independent of
    $\log(p)$, and the bottom thermal damping timescale decreases exponentially
    with $\log(p)$ for $p > 10^5$, staring at 3.3$\tau$; here $\tau$ is the
    rotation period of the planet.  In b), neither type of drag is applied in
    the bottom region.  When both drags are applied, $\bar{u}$ is essentially
    same, irrespective of the initial jet direction; cf. East and West frames in
    a).  However, such a drag is not physical.  When both drags are {\it not}
    applied, $\bar{u}$ is strongly dependent on the initial jet direction---even
    when the  setup permits a supersonic equatorial jet, which is also not
    physical in the simulations [cf. East and West frames in b)].  [from
    Figure~1, \citet{Choetal15}].}
  \label{exop_jet_profile}
\end{figure}

The apparent dichotomy is partly due to the use of different initial, boundary,
and forcing/drag conditions---as demonstrated by \citet{Choetal15}.  An example
is given in Figure~\ref{exop_jet_profile}, which shows time-averaged, zonal-mean
zonal velocity $\bar{u}$ from simulations initialized with a 1000\,m\,s$^{-1}$
east or west equatorial jet.  In all the simulations, strong thermal damping is
applied in the upper region.  In the simulations of
Figure~\ref{exop_jet_profile}a, strong momentum drag along with weak thermal
damping are applied in the lower region; and, in the simulations of
Figure~\ref{exop_jet_profile}b, momentum drag and thermal damping are {\em not}
applied in the lower region.  Note that $\bar{u}$ is essentially the same
regardless of the initial jet direction, in the former case; however,  $\bar{u}$
is strongly dependent on the initial jet direction, in the latter case.  Note
also that the behavior in Figure~\ref{exop_jet_profile}a is in good agreement
with many previous studies whereas the behavior in
Figure~\ref{exop_jet_profile}b is not, clearly demonstrating that a source of
disagreement in simulations is the difference in the dissipation condition
employed \citep{ThraCho11,Choetal15}.  Another main source is numerical in
origin, resulting in variations even under exactly matching physical setup
\citep{Polietal14,Choetal15}.

In the simulations of Figure~\ref{exop_jet_profile}, a prograde equatorial jet
appears to be generic---although its morphology and peak amplitude can be
markedly different between different simulations.  \citet{ShowPolv11} address
the aforementioned category~{\it i}) behavior, focusing on the generation of the
equatorial jet.  They argue---following, e.g., \citet{Wuetal01}---that the jets
result from the interaction of the mean flow with standing Rossby waves induced
by differential (day--night in the exoplanet case) thermal forcing.  In the
proposed scenario, Rossby waves develop phase tilts that pump eastward momentum
from high latitudes to the equator, inducing equatorial superrotation.  An
analytic theory is presented to demonstrate this mechanism as well as a
two-dimensional (shallow-water) model ``support'' for the 3D numerical
simulation results.

However, several caveats should be noted.  First, the shallow-water model does
not cover the entire range of the parameter regime (e.g., supersonic and layer
inhomogeneity) of the category, as have been pointed out explicitly in
\citet{Choetal08}: the shallow-water model is valid, for example, only under
incompressible (i.e., subsonic) condition and cannot be directly compared with
3D simulations exhibiting flows at finite Mach numbers.  Second, a potentially
significant term is included in the momentum equation of the shallow-water model
in \citet{ShowPolv11} to represent the effects of momentum transport between
layers, a 3D effect; but, the representation, which is used for terrestrial
condition, is not valid.  Third, the dependence on the background flow, which
could be crucial, but is unconstrained by theory or observations, is also not
taken into account in the modeling.  Finally, the analytical model shows a weak
symmetry breaking, leading to superrotation only in the (longitudinal) average,
rather than uniformly over the entire equator (as in the 3D simulations): there
is a considerable leap from the shallow-water scenario to the fast multiple
kilometer per second equatorial jets seen in the 3D simulations.

\begin{figure*} 
  \figurebox{33pc}{}{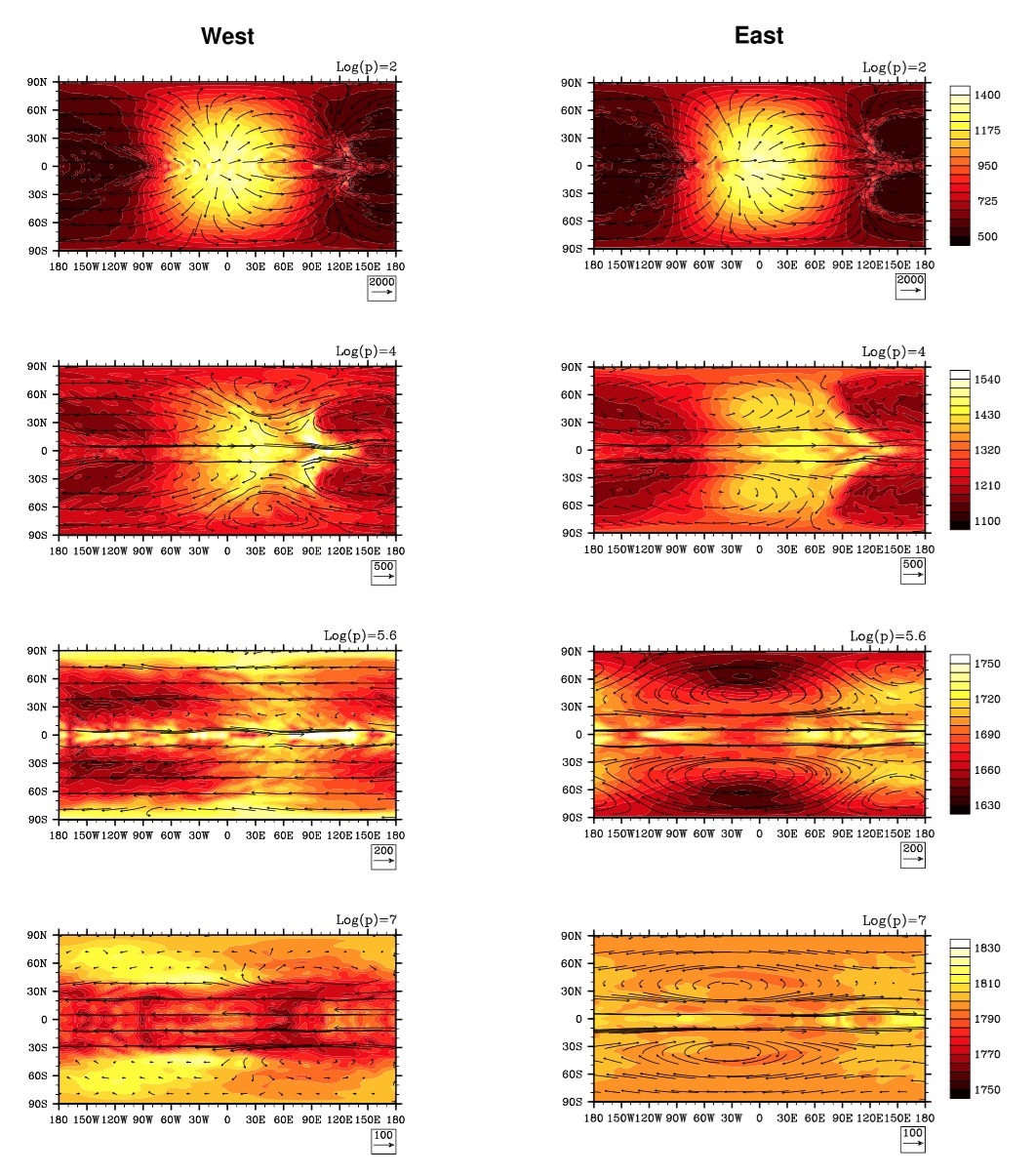}
  \caption{Instantaneous longitude--latitude distributions of the
    flow~(m~s$^{-1}$) and temperature (K) fields at
    $\log(p) = \{2.0, 4.0, 5.6, 7.0\}$, where $p$ is the pressure in Pa, from
    the simulations in Figure~\ref{exop_jet_profile}\,b.  The reference flow
    vectors are shown at the bottom right in each panel, and the temperature
    ranges for each row are shown at the right; the projection is
    cylinderical-equidistant and centered on the substellar point,
    (longitude,\,latitude) = (0,\,0).  Both the flow and temperature
    distributions are complex, with multiple irregularly-shaped hot (cold)
    regions which are often situated away from the substellar (antistellar)
    point.  In the two simulations, both fields are markedly different---except
    near the top of the domain, as expected.  The vertical temperature gradient
    fields in the two simulations are also different, which would lead to
    different emergent heat flux distributions.  [from Figure~3,
    \citet{Choetal15}].}
  \label{exop_temp}
\end{figure*}

Note that bottom drag employed in simulations of Figure~\ref{exop_jet_profile}a
is physically unrealistic.  When such a drag is not used, the difference in the
velocity maps is marked---even when the data is spatiotemporally averaged
(Figure~\ref{exop_jet_profile}b).  When not averaged, the flow and temperature
field can be highly variable, especially in the middle and lower regions of the
domain \citep{ThraCho10,Choetal15}.  This is illustrated in
Figure~\ref{exop_temp}, which shows instantaneous latitude--longitude
distributions of the flow ($\bfv$) and temperature ($T$) fields at four
different pressure ($p$) levels.  The two distributions in both (West/East)
simulations are complex, with multiple irregular hot (cold) regions often
situated away from the substellar (antistellar) point.  Comparing the two
simulations, the instantaneous fields are significantly different, except at
low~$p$ level---as expected, given the very short radiative cooling time there:
the cooling time is proportional to~$p\,/\,T^4$, when the temperature
perturbation of the region associated with the cooling time is small compared to
$T$.  Note also that the vertical temperature gradient fields in the two
simulations are different, which would lead to different emergent heat flux
distributions over the globe as well as a modification of the cooling time
\citep[e.g.,][]{Choetal08}.

\adjustfigure{140pt}

In their study, \citet{KomaShow16} focus on explaining the trend of day-night
temperature contrast on hot-Jupiters with scaling arguments and MITgcm
simulations in the cubed-sphere grid configuration.  They report that the
full-phase infrared light curves of low-eccentricity hot Jupiters show a trend
of increasing dayside--nightside brightness temperature difference with
increasing equilibrium temperature.  In the simulations, they use the Newtonian
cooling scheme as well as a linear frictional drag with two components---a
``basal'' [{\it sic}] drag and a crude parameterization of Lorentz force drag,
following \citet{Pernetal10}; here the basal drag is strongest at bottom of the
computational domain and linearly decreases upward and the ``Lorentz drag'',
applied at the bottom, is spatially constant but varies between different
simulations.  The employed simple drag representation for the species Lorentz
drag has been criticized by \citet{Kosketal14} and \citet{Choetal15} because of
its lack of physical realism for both the upper and lower regions of the modeled
domain.

\begin{figure*} 
  \figurebox{41pc}{24pc}{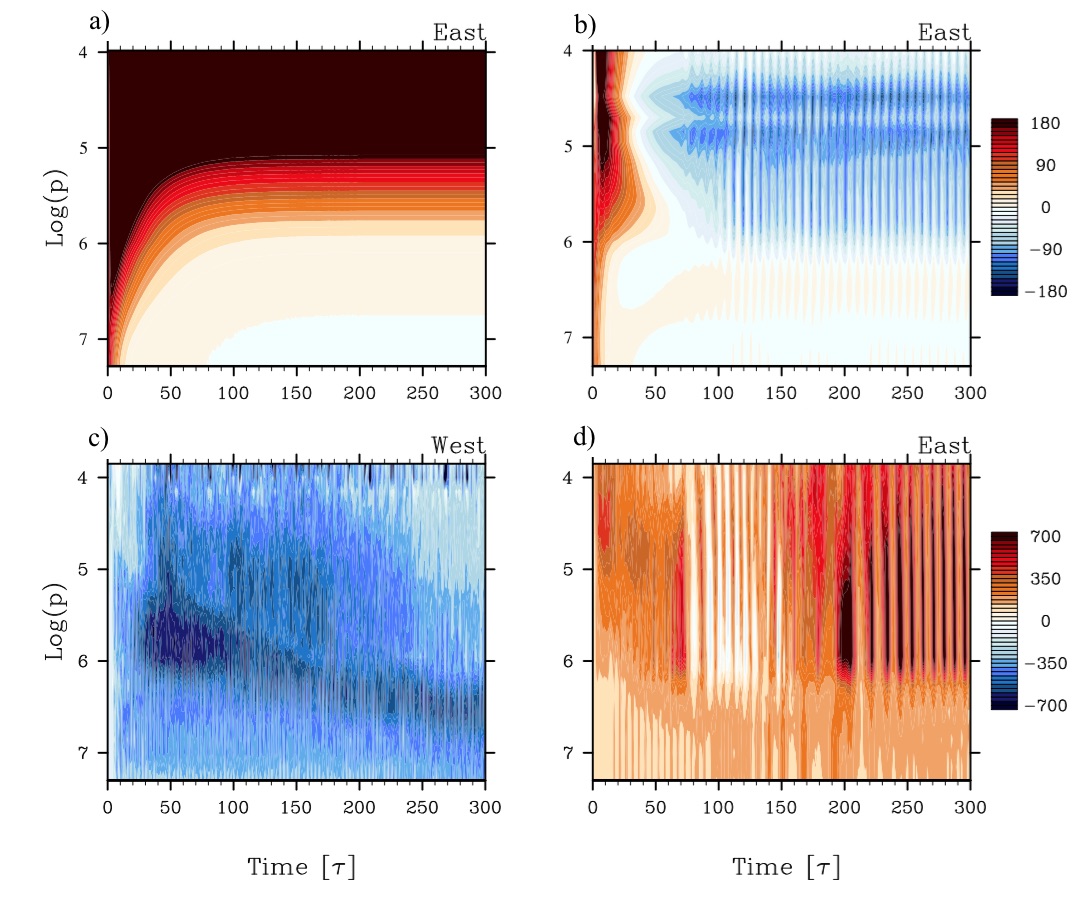}
  \caption{Hovm\"{o}ller plots of instantaneous zonal velocity
    $u(t,p)|_{(\varphi,\vartheta)=(0,30)}$ for $t = [0,300]\,\tau$ and
    $\log(p)=[4.0,7.3]$; here $\varphi$ is the longitude, $\vartheta$ is the
    latitude, $\tau$ is the rotation period, and $p$ is the pressure in Pa.  The
    color bars indicate the velocity (m\,s$^{-1}$) for each row.  Initial jet
    amplitudes and directions (latter indicated above each plot) are as in
    Figure~\ref{exop_jet_profile}.  In a) and b), the resolution is T21L1000
    (i.e., max degree and order of 21 each and 1000 vertical layers) and the
    flow at the bottom is strongly Rayleigh dragged; the vertical levels are
    equally spaced in $\log(p)$ and $p$ for simulations in a) and b),
    respectively.  In c) and d), the resolution is T85L40 and the flow is {\it
      not} dragged near the bottom; the vertical levels are equally spaced in
    $\log(p)$.  In b), c), and d), vertically propagating planetary waves power
    variability, if the vertical resolution near the bottom is increased [cf. a)
    and b)], Rayleigh drag is not applied near the bottom of the domain [cf. a)
    and d)], or sensitivity to initial condition is present [cf. c) and d)].}
  \label{exop_wmfi}
\end{figure*}

\citet{KomaShow16} also argue that the longitudinal propagation of waves
mediates dayside--nightside temperature differences in hot-Jupiter
atmospheres.  Indeed, wave heating and cooling mechanisms have been discussed
explicitly by \citet{WatCho10} for gravity waves and \citet{Choetal15} for
planetary waves; the latter type of waves have been discussed earlier by
\citet{Choetal03}.  Note that both types of waves can be damped in hot-Jupiter
atmospheres by radiative cooling, saturation, and encounters with critical
layers---effecting communication over long distances (``teleconnection''), in
both the lateral and vertical directions.  An example of the propagation and
interaction of planetary waves can be seen in Figure~\ref{exop_wmfi}, in which
upwardly propagating planetary waves strongly modify the background jet (and
associated temperature) structure.  In the figure, propagating waves power
variability, if 1) vertical resolution near the bottom is increased
[cf. Figure~\ref{exop_wmfi}a and Figure~\ref{exop_wmfi}b], 2) linear Rayleigh
drag is not applied near the bottom of the domain [cf. Figure~\ref{exop_wmfi}a
and Figure~\ref{exop_wmfi}d], or 3) sensitivity to initial condition is present
[cf. Figure~\ref{exop_wmfi}c and Figure~\ref{exop_wmfi}d].

\citet{Parmetal16} use the thermal structure from MITgcm simulations to
determine the expected cloud distribution and Kepler light-curves of
hot-Jupiters.  Post-processing the simulation data with plane-parallel radiative
transfer and ``equilibrium cloud'' models, they report that the change from an
optical light-curve dominated by thermal emission to one dominated by scattering
(reflection) naturally explains the observed trend from positive to negative
offset of the hot spot location.  They speculate that, for the cool planets, the
presence of an asymmetry in the Kepler light curve is indicative of the cloud
composition, because each cloud species can produce an offset only over a narrow
range of effective temperatures.  They also add that the cloud composition of
hot-Jupiters likely varies with equilibrium temperature, suggesting that a
transition occurs between silicate and manganese sulfide clouds at a temperature
near 1600\,K, analogous to the L/T transition on brown dwarfs: the cold trapping
of cloud species below the photosphere naturally produces such a transition and
predicts similar transitions for other condensates, including TiO.  Of course,
all of these results---as well as those below---depends on the accuracy of the
jet and thermal structures obtained in the GCM simulations.

\citet{Kataetal16} present results from a MITgcm simulation study of nine
hot-Jupiters that make up a large transmission spectral survey using the Hubble
and Spitzer Space Telescopes.  These observations exhibit a range of spectral
behavior over optical and infrared wavelengths, suggesting perhaps diverse cloud
and haze distributions.  By utilizing the specific system parameters for each
planet, they explore the parameter-space spanned by planet radius, surface
gravity, orbital period, and equilibrium temperature.  In this study, they show
that their model ``grid'' recovers trends shown in traditional parametric
studies of hot-Jupiters, particularly equatorial superrotation and increased
day-night temperature contrast with increasing equilibrium temperature.  They
report that spatial temperature variations---particularly between the dayside
and nightside and west and east terminators---can vary by hundreds of degrees,
which may imply large variations in Na, K, CO, and CH4 abundances in those
regions.  They also compare theoretical emission spectra generated from their
models to available Spitzer eclipse depths for each planet and find that the
outputs from their cloud-free, solar-metallicity models generally provide a good
match to many of the datasets.

\citet{Maynetal14,Maynetal17} have adapted the Unified Model (UM) and use it to
study {\it HD209458b}.  They compare their results with test cases,
\citet{Showetal09}, and available observations.  They also focus on the
robustness and evolution of a superrotating equatorial jet and its interaction
with the deep atmosphere.  The UM solves the full 3D Navier-Stokes equations
with a height-varying gravity, appropriate for deep atmospheres; these equations
are valid for non-hydrostatic atmospheres.  In their study, the occurrence of a
superrotating, supersonic equatorial jet is robust to changes in various
parameters they consider, and over long timescales---even in the absence of
strong inner or bottom boundary drag, similar to what has been reported in
\citet{Choetal15}.  Note that \citet{Choetal15} have argued that such a jet is
not physical in their simulations, since supersonic jets are produced under
hydrostatic conditions with rigid (free-slip) top and bottom boundary
conditions.  It is also worth noting that shocks are not resolved in simulations
in any of the studies.  As an aside, Mayne et al. (2017) also find that the jet
amplitude is diminished when the equator-to-pole temperature gradient in the
deep atmosphere is forced over long timescales, showing dependence on the
details of the setup.

\citet{Amunetal16} extend the model of \citet{Maynetal14} by incorporating a
radiation scheme based on the two-stream approximation and correlated-{\it k}
method with opacities from the {\it ExoMol} database \citep{TennYurc12} and
compare their results with observations and \citet{Showetal09}.  In their study,
\citet{Amunetal16} find a reasonable agreement between observations and both
their day-side emission and hot spot offset.  But, they report that their night
side emission is too large.  In addition, although their results are
qualitatively similar to those of \citet{Showetal09}, they report several
quantitative differences: their simulations show significant variation in the
position of the hottest part of the atmosphere with pressure, as shown in Cho et
al. (2015).  This is in contrast to the ``significant vertical coherency''
reported by \citet{Showetal09}.  In addition, they also find significant
quantitative differences in calculated synthetic observations.  In
\citet{Amunetal16} the temperature-pressure profile in their deep atmosphere
continues to evolve, without reaching equilibration.

Mendon\c{c}a et al. (2016) introduce and show benchmark tests for their new GCM,
THOR, that solves the 3D non-hydrostatic ``Euler equations''.  Their model uses
an icosahedral grid to address the pole problem (as with the cubed-sphere grid
for MITgcm) and  a tunable explicit dissipation ``spring dynamics'' scheme,
chosen to ensure stable model integration.  Hence, the strength of dissipation
does not change with latitude or longitude.  Note that, although inviscid
equations are solved, substantial dissipation of numerical origin is still
present.  In addition, split-explicit method is used for the time stepping,
together with a horizontally explicit and vertically implicit integration.  The
model is designed to run on graphical processing units (GPUs) and is a part of
the open-source Exoclimes Simulation Platform.  The developers have validated
their code with the Held-Suarez test case for Earth-like conditions and
qualitatively reproduced the results of \citet{MenoRaus09} and
\citet{Hengetal11} for hot-Jupiter-like conditions.

\adjustfigure{170pt}

\citet{Frometal16} use RAMSES, a finite-volume shock-capturing code, to also
solve the compressible ``Euler equations'' in the {\it beta-plane}; the
beta-plane is a tangent plane approximation, retaining the Corliois term
contribution to first order at a specified latitude.  These investigators focus
on the effects of compressibility, shear-driven instabilities, and shocks in
hot-Jupiter atmospheres.  At low resolution, they observe a steady supersonic
equatorial jet of a few kilometers per second that does not display shocks.  At
higher resolution, they show that the equatorial jet is unstable to both a
barotropic Kelvin-Helmholtz instability and a vertical shear instability.  The
jet zonal mean velocity displays regular oscillations with a typical timescale
of a few days and a significant amplitude of about 15\% of the jet
velocity.  They also report finding compelling evidence for the development of a
vertical shear instability at pressure levels of a few bars, which seems to be
responsible for an increased downward kinetic energy flux that significantly
affects the temperature of the deep atmosphere and appears to act as a form of
drag on the equatorial jet.  This instability also creates velocity fluctuations
that propagate upward and steepen into weak shocks at pressure levels of a few
mbars (n.b., 1\,bar = $10^5$\,Pa).  They conclude that hot-Jupiter equatorial
jets are potentially unstable to both a barotropic Kelvin-Helmholtz instability
and a vertical shear instability.

Prior to this study, \citet{PoliCho12} solve the hydrostatic 3D equations on the
full sphere and demonstrate the possibility of baroclinic instability in
hot-Jupiter-like conditions; the instability is sensitive to the morphology of
the jet and the temperature distribution.  In their study, when it occurs, the
instability evolves concomitantly with a barotropic instability.  The
instability naturally transports kinetic energy downward, as part of the
equilibration process.  Additionally, they show that, for a high-speed subsonic
{\it equatorial} jet, instability occurs at the flanks (but not at its
core), when adequate resolution is present---as confirmed by \citet{Frometal16},
solving the non-hydrostatic 3D equations on the beta-plane.

GCMs solving the full Navier-Stokes equations have also been directly coupled
with detailed cloud models.  \citet{Helletal16} extend the model used by
\citet{Dobbetal10} and \citet{DobbAgol13} and focus on aspects of cloud
distribution on {\it HD189733b} and {\it HD209458b}.  \citet{Helletal16} report
that, in their study, both planets are covered in mineral clouds throughout the
entire model domain---independently of differences in hydrodynamic
models.  Further, the clouds are chemically complex, composed of mineral
particles that have a height-dependent material composition and size: therefore,
single values to characterize metallicity and C/O ratio are not valid.  They
also argue that their results concerning the presence and location of water in
relation to the clouds explain some of the observed difference between the two
planets and that obscuring clouds exist high in the atmosphere of {\it
  HD209458b}, but much deeper in {\it HD189733b}.

However, the above interpretation does not uniquely or wholly fit the currently
available transit spectra.  For example, in the case of {\it HD209458b}, a clear
atmosphere provides a satisfactory fit to the transit data; see, e.g., Figure~7
of \citet{Lavvetal14}.  In particular, the Na~D line wing and the 1.4~micron
water band show up roughly at the same level as the short wavelength slope,
indicating that these features probe a similar pressure level.  The revised data
from \citet{Singetal16} also presently show a possible K line wing.  On {\it
  HD189733b}, fitting the HST transit spectrum with a clear atmosphere is
difficult; see, e.g., Figure~7 of \citet{LavvKosk17}.  There are no Na~D or
K~line wings and the short wavelength slope lies well above the 1.4~micron water
feature.  This configuration requires the presence of a high altitude absorber
with a small particle size.  Indeed, high altitude haze composed of soots can
form on {\it HD189733b} with abundances sufficient to explain the alleged
transit spectrum.

\citet{Leeetal16} also use an extended model of \citet{Dobbetal10} and
\citet{DobbAgol13}, focusing on clouds formation under 3D dynamics for {\it
  HD189733b}.  The simulation includes the feedback effects of cloud advection
and settling, gas phase element advection and depletion/replenishment, and the
radiative effects of cloud opacity.  The cloud particles are modelled as a mix
of mineral materials, which change in size and composition as they travel
through atmospheric thermo-chemical environments.  All local cloud properties
such as number density, grain size, and material composition are
time-dependent.  Gas phase element depletion as a result of cloud formation is
included in the model.  In situ effective medium theory and Mie theory are
applied to calculate the wavelength-dependent opacity of the cloud component.

In their study, \citet{Leeetal16} find that the mean cloud particle sizes are
typically sub-micron (0.01--0.5 $\mu$m) at pressures less than 1 bar, with
hotter equatorial regions containing the smallest grains.  Denser cloud
structures occur near the terminators and in the deeper regions ($>$\,1
bar).  Silicate materials, such as MgSiO3(s), are found to be abundant at
mid--high latitudes, while TiO2(s) and SiO2(s) dominate the equatorial
regions.  Elements involved in the cloud formation can be depleted by several
orders of magnitude.  The interplay between radiative-hydrodynamics and cloud
kinetics leads to an inhomogeneous, wavelength dependent opacity cloud structure
with properties differing in longitude, latitude, and depth.  This suggests that
transit spectroscopy would sample a variety of cloud particles properties (e.g.,
sizes, composition, and densities).  Presumably, such modeling could in
principle help to elucidate or constrain information about jets.

Several studies have appeared in the literature very recently and we briefly
mention them here.  \citet{Tremetal17}, based on a two-dimensional steady-state
atmospheric circulation model, model the advection of the potential temperature
$\Theta$ due to mass and longitudinal momentum conservation; here
$\Theta = T(p_{\mbox{\tiny R}}/p)^\kappa$, where $p_{\mbox{\tiny R}}$ is a
reference pressure (often set to 1 bar) and $\kappa$ is the adiabatic
index.  They argue that longitudinal--vertical mixing imply a larger radius for
the planet, reproducing the observed radius of {\it
  HD209458b}.  \citet{RogeMcEl17} show that a dynamo can be maintained in the
atmosphere of a hot-Jupiter by conductivity variations arising from strong
asymmetric heating from the planets’ host star, independently of the deep-seated
dynamo in the planet.  They argue that the presence of a dynamo significantly
increases the magnetic field strength on the surface.  \citet{ZhanShow17}
investigate the effects of atmospheric bulk compositions on temperature and wind
distributions for tidally locked sub-Jupiter-sized planets, using the
MITgcm.  \citet{PennVall17}, using a shallow-water model with time-dependent
forcing, confirm the inhomogeneous shallow-water equations study by
\citet{Choetal08} that the peak of an exoplanet thermal phase curve is, in
general, offset from the secondary eclipse when the planet is rotating.  They
also consider the inverse problem of constraining planetary rotation rate, a
critical parameter for dynamics, from an observed phase curve.

\subsection{Some Critical Issues}\label{exo_iss}

As mentioned, current simulations are in agreement about the small number
(usually three) of broad jets expected on synchronized giant
exoplanets.  However, it is not clear that the agreement is entirely
significant, given that a number of modeling issues remain to be addressed and
resolved.  The situation should certainly improve as more repeated observations
of the same planets are made, which could provide better constraints.  Crucial
scales and parameters---such as the Rossby deformation scale $L_R$, Rhines scale
$L_\beta$, and plasma beta parameter $\beta_p$ \citep[see, e.g.,][]{Cho08}---are
still poorly known, and pose a great challenge for simulation as well as
overall understanding; here $L_R$ is a measure of the distance traversed by a
propagating gravity wave in one rotation period of the planet $\Omega_p$;
$L_\beta$ is the scale at which nonlinear proceses is balanced by the gradient
of the Coriolis acceleration and can serve as a crude measure of jet width; and,
$\beta_p$ is a measure of the speed of the gravity wave relative to the
Alfv\'{e}n wave.  Paradoxically (perhaps), the jet strength is not proportional
to the separation distance of the planet from its host star \citep{Choetal08},
at least in the Solar System: there is some evidence that the peak strength of
the jets may be inversely proportional to the separation distance and the depth
in the atmosphere where the heat deposition occurs.

Studies already clearly show that improved numerical algorithms and models with
greater resolution and physical complexity are crucially needed for a better
understanding of jets on exoplanets \citep[see,
e.g.,][]{PoliCho12,Polietal14,Choetal15}.  One reason for this is because many
of the planets are in highly {\em a}geostrophic, ``unbalanced'' regimes.  Higher
resolution will be needed to resolve the narrower jets that result from faster
rotation rates expected for synchronized planets closer-in than approximately
0.02\,AU, for example.  Also, small-scale gravity waves would be even more
dominant than in present situations, and would be even more crucial for
accuracy.  Additionally, ideal and non-ideal magnetohydrodynamic (MHD)
simulations will also need to play a greater role, as ionization becomes
stronger---even at depths much greater than previously thought
\citep{Kosketal14,Choetal15}.

To this point, some simulations have employed a linear, Rayleigh drag as a
``crude representation of magnetic drag effects'' that putatively stem from
thermal ionization.  As pointed out in \citet{Choetal15}, there are two major
concerns with this.  First, thermal ionization is insignificant in the modeled
region, as temperature is too low and density is too high.  This is so even
taking into account the low-ionization potential of alkali metals (e.g., K, Na,
Ca) because these are species are not present in abundant amounts, assuming
solar abundances---i.e., $n_{\rm H+} / n_{\rm n}\, \lsim\, 2 \times 10^{−16}$
and $n_{\rm K+} / n_{\rm H+} \approx 3 \times 10^6$, where $n_{\rm x}$ is the
x-specie number density and ``n'' subscript refers to the neutral component.
Secondly, even if the ion-induced drag were significant via a non-thermal
mechanism, for example, it cannot be represented as a simple isotropic
drag-to-rest on the momentum field: ion velocities and the intrinsic field
orientation need to be modelled self-consistently for accurate representation
\citep[e.g.,][]{Kosketal10}.

While modeling work to date has covered parts of the presumably relevant
parameter-space with different types of physical and numerical models, it is
important to note that many non-trivial modeling choices made are often rather
similar between different studies.  For example, when the models are forced with
temperature relaxation, the assumptions about the structure of equilibrium and
initial temperature distributions do not reflect the large uncertainty and range
of conditions that are plausible.  It is not clear if more complex forcing
(e.g., incorporating sophisticated radiative transfer), alleviates this
uncertainty or even worsen the situation.  Likewise, the uncertainty and effects
of deeper layers of the atmospheres has only been addressed to limited
extent. Usually, in the models, the deeper layers contain a large mass of slowly
evolving air that is poorly resolved and almost always assumed to start from
rest. But in reality, this large mass of air could have a variety of dynamic
patterns that could presumably affect the layers above.

Verification of newly developed or adapted codes for highly nonlinear
atmospheric dynamics problems is not a trivial task.  Rigorous testing with
analytical solutions is not possible and equatable comparison with other
published solutions is difficult because they usually show chaotic
time-dependence and because specific assumptions (e.g., on boundary conditions
or neglected terms) are not implemented in different codes or often not even
adequately detailed.  This generates unnecessary confusion.  

A community effort for concerted validation and comparison of the numerous codes
currently in use would be extremely valuable at this time, as shown by
\citet{Polietal14} and \citet{Choetal15}.  These studies clearly show that one
can arrive at erroneous conclusions if simulations from different codes are not
at least qualitatively reproduced with the same setup.  Even then, erroneous
conclusions can still be drawn, as codes which perform well ostensibly in one
region of the physical parameter-space do not perform well in another (or, more
precisely, extended) region.  For example, when pushed to a highly {\em
  a}geostrophic region, numerical accuracy of the code can become seriously
degraded by small-scale, fast oscillations generated in that part of the
parameter-space \citep[e.g.,][]{ThraCho11}.  In addition, the atmosphere can
exhibit multiple equilibrium states or be driven unwittingly to an unphysical
state \citep[][Cho and Thrastarson (in prep.)]{Choetal15}.  In general, the
critical question of ``realistic and accurate'' forcing and initialization---and
their effects---is still unsettled, and the question deserves much more
attention and scrutiny than has generally received thus far.

\section{THE SUN}\label{sun}

We now turn our attention to the Sun.  The Sun is a Main-sequence star.  Such
stars share a number of common dynamical elements, which include turbulent
convection, differential rotation, and magnetic activity.  They ``burn''
hydrogen in their interiors and exhibit a wide range of fluid dynamical
behaviour, under various arrangements of radiative and convective zones.  For
example, in the Sun, $p$--$p$ chain fusion reaction leads to a radiative core
(RC) but the much more rapid C-N-O nuclear burning cycle outside the core gives
rise to a convective envelope outer layer, called the convection zone~(CZ).  The
CZ occupies the 29\% of the Sun's radius ($R_\odot = 6.96\times 10^8$\,m), from
its surface down to $\sim\! 2\!\times\! 10^8$\,m in depth.  On the surface,
vigorous convection, differential rotation (manifested by broad zonal flows) and
magnetic cycles are readily visible
\citep[e.g.,][]{Mies05,Solaetal06,MiesToom09,Prie14}.  Not only are RC and CZ
tightly coupled, the jets are thought to be intimately related to the boundary
processes occurring in them, including the meridional (poleward as well as
toward the rotation axis) flow and entropy gradient---as we discuss below.

\subsection{Observations}\label{sun_obs}

\begin{figure}
  \figurebox{20pc}{}{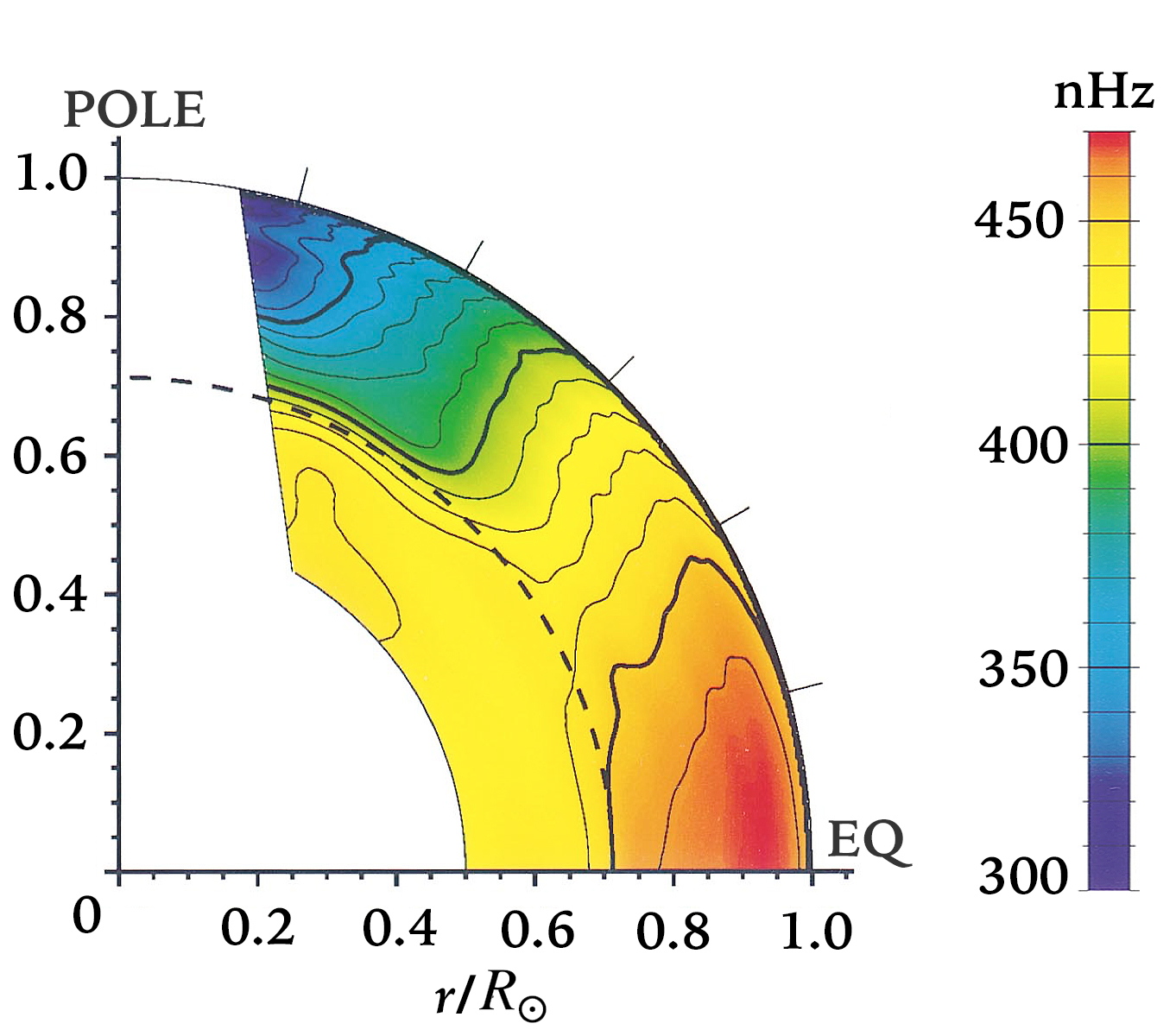}
  \caption{Angular velocity profile $\Omega_\odot(r,\theta)$, where $r$ is the
    radius and $\theta$ is the {\it co}latitude, obtained via helioseismology;
    $R_\odot$ is the radius of the Sun.  The convection zone (CZ) is bounded
    above by a shear layer that extends from $\sim$0.95$R_\odot$ to the surface
    and below by a even stronger shear layer, the tachocline, located near the
    CZ base at $\sim$0.71$R_\odot$; the tachocline straddles the bottom of the
    CZ and the top of the radiative core~(RC).  Note the essentially radial
    profiles of $\Omega_\odot$ in the interior of the CZ: the profile signifies
    nearly radially-barotropic, deep zonal jets (defined with respect to the
    nearly solid-body rotation of the RC).  [from Figure~5b,
    \citet{Schoetal98}].}
  \label{sun_rot_profile}
\end{figure}

Unlike for the stars that host the exoplanets, the surface activity on the Sun
can be observed in exquisite detail.  Remarkably, even its interior flow can be
probed.  This is because of the Sun's proximity.  NASA's Solar Dynamics
Observatory (SDO) mission has already provided a large amount of data on solar
dynamics and magnetic activities, and this data is complemented by the data from
the Solar and Heliospheric Observatory (SOHO) mission as well as by ground-based
observatories, which include the Global Oscillation Network Group (GONG) and the
New Solar Telescope (NST).  Helioseismic soundings using resonant $p$-modes
(acoustic waves) have produced the angular velocity profile,
$\Omega_\odot = \Omega_\odot(r,\theta)$, where $r$ is the radial distance from
the center and $\theta$ is the {\em co}latitude, for most of the Sun's interior
\citep[e.g.,][]{GougToom91,Chri-Dals02}; note that $\Omega_\odot$ is not a
constant here, in contrast to $\Omega_p$.  The $\Omega_\odot$ profile is shown
in Figure~\ref{sun_rot_profile}.

The Sun's visible surface has long been known to rotate differentially, with a
rotation period of $\sim$25\,days near the equator and $\sim$33\,days at
high-latitude regions.  This leads to a roughly ``three zonal jet'' structure,
pole-to-pole.  These jets are referenced against the quasi-uniform rotation of
the RC.  The $\Omega_\odot$ profile obtained from the soundings clearly shows
the deep structure and the remarkable feature that the rotation contours are
roughly constant on conic surfaces (rather than on cylinderical surfaces).  Note
also that the CZ is bounded by a shear layer near the surface and another, very
strong shear layer---called the {\it tachocline} \citep{SpieZahn92}---that span
across the bottom of the CZ and the top of the RC.

Doppler measurements of the photosphere and local helioseismic inversions show a
poleward meridional flow in the surface layers
\citep{Hath96,Habeetal02,ZhaoKoso04}.  Such an {\it a}zonal flow, which is
nevertheless closely tied with the mean zonal flow, may be driven by the
so-called ``gyroscopic pumping'' mechanism \citep{McIn98}.  In this mechanism, a
negative radial shear (i.e., outward reduction) of $\Omega_\odot$ near the
surface suggests a local retrograde (westward) mean flow, which is driven by the
divergence of Reynolds stresses associated with super-granulation.  In this way,
poleward circulation may be considered as a ``boundary effect'' that occurs near
the surface.  Although helioseismic inversions so far suggest that the shear
layer penetrates too deeply into the CZ for this mechanism to operate
effectively \citep{Gileetal97,BrauFan98}, the inversions are currently not
sensitive enough to definitively rule it out.

At the lower boundary of the CZ, if the tachocline is in thermal wind balance
\citep[see, e.g.,][]{Tass00}, then the strong radial rotational shear from
helioseismic inversions entail the presence of latitudinal thermal
gradients.  This is expressed by the relation,
\begin{eqnarray}\label{twbe}
  \qquad \qquad \qquad \quad 
  \bfOmega_\odot\!\cdot\!\nabla\,\overline{v}_\varphi \ = \
  \frac{g}{2c_p\,r}\,\frac{\del\overline{s}}{\del\theta}\, , 
\end{eqnarray}
where $\bfOmega_\odot = \Omega_\odot\,\mbox{\bf e}_z$ is the rotation vector,
$v_\varphi$ is the longitudinal (eastward) velocity component and the overbar
denotes an average over the longitude $\varphi$ and time $t$, $g$ is the
gravitational acceleration, $c_p$ is the specific heat per unit mass at constant
pressure, and $s$ is the specific entropy per unit mass
\citep[e.g.,][]{BrunToom02,Durn99,Ellietal00,KitcRudi95,Mies05}.  Note that
Equation~\ref{twbe} is valid for an ideal gas in hydrostatic balance with nearly
adiabatic background and with $R_o \ll 1$.

The existence of the tachocline, and its associated overshoot region, may have a
profound effect on the dynamo action and mean flow in the Sun---for example, in
promoting the generation of strong toroidal fields, cyclic variability, and
zonal jets.  Turbulent motions in the convection zone expel magnetic fields
toward the boundaries, and the asymmetry between upward and downward flows gives
rise to a systematic pumping of mean and fluctuating fields downward
\citep{Tobietal01,ZiegRudi03}.  The stable stratification of the overshoot
region helps to suppress magnetic buoyancy instabilities; and, the rotational
shear of the tachocline amplifies mean toroidal fields, which could enhance the
dissipation of small-scale fields through the distortion and fragmentation of
closed magnetic loops.  The latter is related to the rotational smoothing
process, in which opposite field polarities are brought together by vertical
shear and are dissipated on a timescale of
$\sim\![\lambda^2/(\eta \delta^2)]^{1/3}$ \citep{Spru99}; here $\lambda$ is the
longitudinal wavelength, $\eta$ is the magnetic diffusivity, and
$\delta = {\rm d}U / {\rm d}r$ is the radial shear with $U$ the characteristic
speed.

Although high-precision oscillation measurements have greatly advanced our
understanding of the dynamical structure of the Sun's surface and interior,
several important issues remain to be resolved.  For example, there is a
discrepancy between the spectroscopically-determined abundance of heavy elements
on the surface and the abundance deduced from global helioseismology data (in
conjunction with numerical modeling), indicating that understanding of the basic
physics of the interior is still incomplete.  This includes fundamental issues
related to the properties of non-ideal plasma, which affect the accuracy of
abundance estimates in helioseismology.  In addition, it is still not known
whether the inner, energy-generating part of the RC rotates faster or slower
than the outer part.  This is of fundamental importance for understanding the
overall formation and evolution of the Sun---including the jets.

\subsection{Jets in Simulations}\label{sun_sim}

Numerical simulations of 3D convection in a rotating spherical annulus have
become quite advanced.  In these simulations, the large-scale flow is in a state
close to a ``geostrophic balance''  \citep[see, e.g.,][]{Tass00,Chri02}, if
$Ro \ll 1$.  In this situation, Equation~\ref{twbe} reduces to the
Taylor-Proudman theorem, in which zonal velocity (specifically, $\Omega_\odot$)
contours are cylindrical and aligned with the rotation direction
$\bfOmega_\odot$.  In a less rapidly-rotating annulus (which is more appropriate
for the Sun), a non-cylindrical rotation component is exhibited and maintained
by the baroclinic forcing term on the right-hand side of
Equation~\ref{twbe}.  The baroclinicity is such that the rotation profile is
more conical, with $\Omega_\odot$ decreasing toward the poles.  This would
entail a poleward entropy gradient (that is, $\del\overline{s} / \del\theta < 0$
in the northern hemisphere).  However, it is important to note that
Equation~\ref{twbe} is not valid for a slowly rotating (i.e., $R_o \gsim 1$)
object.

\begin{figure*} 
  \figurebox{42pc}{}{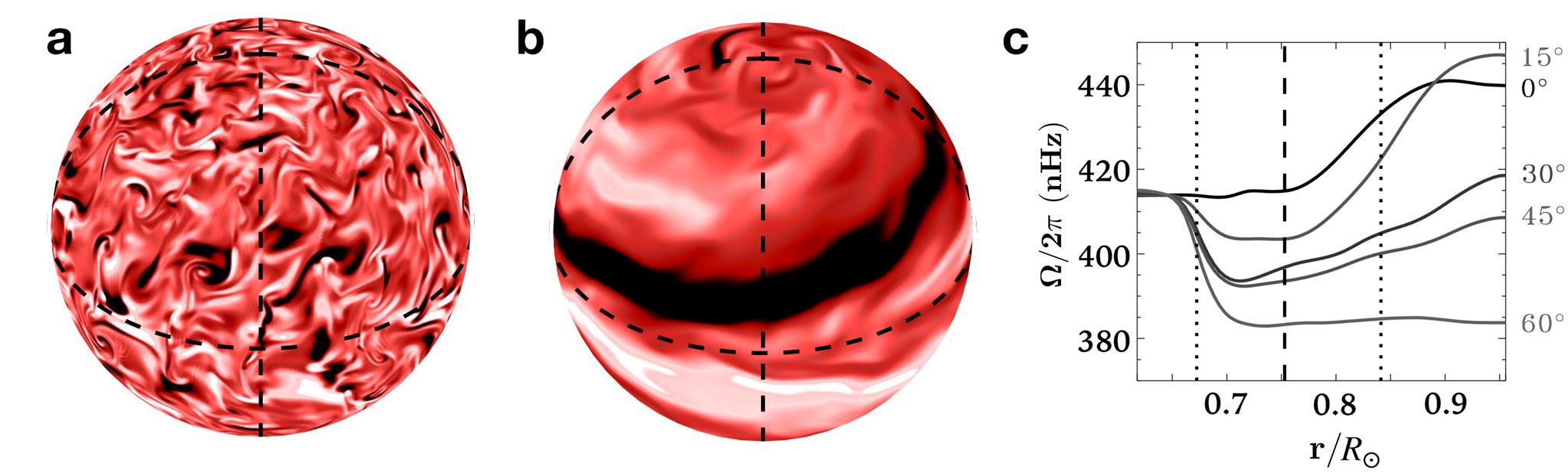}
  \caption{Pumping, organization, and amplification of the magnetic fields in a
    convective dynamo simulation which incorporates a tachocline-like rotational
    shear layer at the bottom of the CZ.  Instantaneous longitudinal magnetic
    field component B$\varphi$ is shown in orthographic projection for a) the
    mid-CZ and b) the tachocline, with white/pink and red/black denoting
    eastward and westward fields, respectively.  The mean angular velocity
    versus radius is shown c) for selected latitudes.  The vertical dashed line
    denotes the ``base'' of the convection zone, and the dotted lines denote the
    depths depicted in a) and~b).  [from Figure~5, \citet{Mies08}].}
  \label{sun_conv_sim}
\end{figure*}

On the other hand, although the Sun is often thought of as a slowly rotating
star, the rotation is rapid enough that the meridional acceleration of zonal
shear flows by the Coriolis force can exceed the convective momentum transport
by the Reynolds stresses in the meridional plane.  This is because {\it locally}
$R_o$ can be much less than unity---for example, if the speed and length scales
based on the Reynolds stress (and $\Omega_\odot$ of the RC) are used.  The
smallness of $R_o$ for such structures appear to be well-supported by 3D global
convection simulations \citep[e.g.,][]{BrunToom02,Ellietal00}.  This may also be
of importance for gaseous exoplanets.  Note that, for the Earth, $R_o$ for
convective scales tends to be large, in contrast to what is seen here.

Global solar convection simulations that incorporate a tachocline of rotational
shear do indeed exhibit pumping, organization, and amplification of toroidal
fields \citep{Browetal06}.  This can be seen in Figure~\ref{sun_conv_sim}.  In
the figure, the CZ is dominated by fluctuating, non-axisymmetric fields that
account for approximately 95\% of the total magnetic energy
(Figure~\ref{sun_conv_sim}a).  In contrast, the field in the tachocline is
dominated by oppositely directed, axisymmetric toroidal bands in the northern
and southern hemispheres (Figure~\ref{sun_conv_sim}b) that account for more than
60\% of the magnetic energy and that which have persisted for a simulated
duration of 17 (Earth) years, apart from a brief 105-day interval of symmetric
parity \citep{Browetal07}.  The mean poloidal field is similarly more ordered,
with a stronger dipole moment and less frequent reversals.

In general, rotational shear can be established through the advection of angular
momentum by meridional circulation, as well as through convective Reynolds
stresses.  In a statistically steady state, these must be balanced so that
\begin{eqnarray}\label{sfe}
  \qquad \quad
  \nabla\cdot\Big[\,\rho_0\,\Big(\overline{\bm{v}_m}\,\mathscr{L} - 
  \overline{v^\prime_\varphi \bm{v}^\prime_m}\, r\sin\theta\Big)\,\Big]\ =\ 0\, ,
\end{eqnarray}
where $\bm{v}_m$ is the meridional velocity component, $\rho_0 = \rho_0(r)$ is
the density averaged over horizontal surfaces and time,
$\mathscr{L} = r \sin\theta\,(\,\Omega\,r \sin\theta + \overline{v_\varphi}\,)$
is the specific angular momentum \citep{Tass00}, and primes indicate small
perturbations about the mean.  More broadly, Equation~\ref{sfe} generalizes to
\begin{eqnarray}\label{fde}
  \qquad \qquad \qquad 
  \rho_0\,\overline{\bm{v}_m} \cdot \nabla \mathscr{L}\ 
  =\ -\nabla \cdot {\bf F}\, ,
\end{eqnarray} 
where {\bf F} is an angular momentum flux that includes contributions from 
Reynolds and Maxwell stresses as well as viscous diffusion.  Note that the
latter is negligible in stars, but not in numerical simulations.

According to Equation~\ref{fde}, an acceleration of the jet
$\overline{v_\varphi}$, powered by the divergence of ${\bf F}$, induces a
meridional flow across constant surfaces of $\mathscr{L}$.  This is effected
through the Coriolis acceleration of $\overline{v_\varphi}$ and may also involve
a contribution from the thermal wind, which satisfies
Equation~\ref{twbe}.  Because $\nabla\mathscr{L}$ is directed cylindrically
outward, away from the rotation axis, retrograde and prograde forcing induce
circulations toward and away from the rotation axis, respectively: this is the
gyroscopic pumping mechanism discussed above.  But, note that cylindrically
inward gradients satisfy the necessary condition for instability according to
the Rayleigh criterion \citep[see, e.g.,][and references therein]{Tass00}. In
the gyroscopic pumping, circulations are set up in response to the convective
Reynolds stress and the rotational shear, which satisfy the dynamical balance of
Equations~\ref{twbe} and~\ref{fde}---a delicate balance between large
forces.  Hence, the meridional circulation in numerical simulations typically
exhibits very large fluctuations ($\sim$300\%) about its temporal mean
\citep{MiesToom09}.  This is probably not realistic.

\begin{figure*} 
  \figurebox{42pc}{}{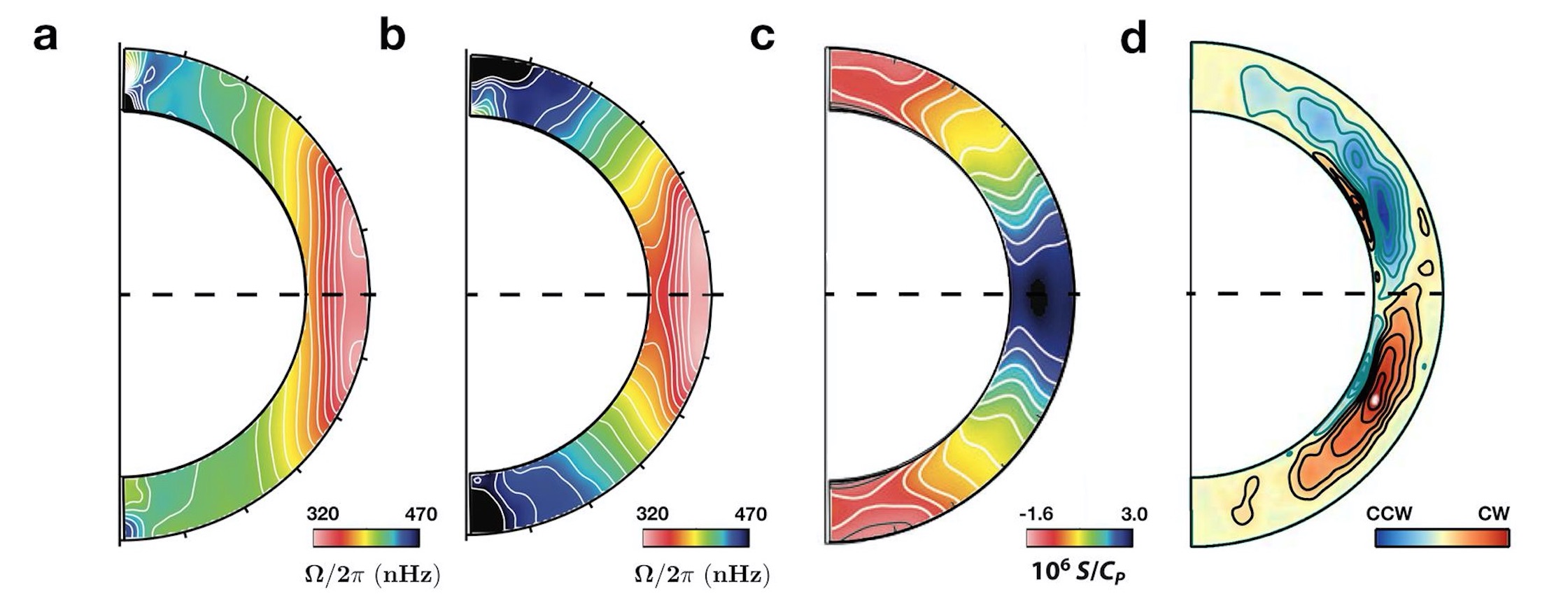}
  \caption{a,b)~Mean angular velocity $\Omega_\odot/2\pi$, c)~specific entropy
    perturbation, and d)~meridional circulation in simulations of solar
    convection, averaged over longitude $\varphi$ and time $t$.  The simulation
    in a) has a uniform specific entropy specified at the lower boundary, and
    the simulation in b) has a poleward latitudinal entropy gradient imposed on
    the lower boundary.  The simulation in c) correspond to the same simulation
    and time interval as in b).  Contour intervals in~a) and~b) are 10\,nHz and
    in~d) represent streamlines of the mass flux, with red and blue denoting
    clockwise~(CW) and counterclockwise~(CCW) circulation, respectively.  [from
    Figure~6a--c, \citet{Miesetal08} and Figure~6d, \citet{Miesetal06}].}
  \label{sun_ang_vel}
\end{figure*}

Baroclinicity, which could aid in establishing a non-cylindrical rotation
component, does arise in global convection simulations of the Sun.  As noted,
this is due to the influence of rotation on the convective heat flux, which
tends to establish a poleward entropy gradient and non-cylindrical
$\Omega_\odot$ profiles \citep{BrunToom02,Ellietal00, Miesetal00}.  However,
\citet{Miesetal06} suggest that thermal coupling to the tachocline may also play
a significant role in establishing a non-cylindrical profile.  Here the
sub-adiabatic stratification of the lower tachocline is essential for ensuring
that a Sun-like rotational shear can generate a poleward entropy gradient via
meridional circulation.  Ultimately, the profile obtained depends on a complex
interaction between convection, differential rotation, meridional circulation,
and thermal stratification.  This is illustrated in Figure~\ref{sun_ang_vel}.

Reynolds stresses maintain a significant differential rotation but angular
velocity contours are cylindrical in accordance with the Taylor-Proudman
theorem, when a uniform entropy is imposed at the lower boundary
(Figure~\ref{sun_ang_vel}a).  On the other hand, a conical rotation profile is
obtained (Figure~\ref{sun_ang_vel}b), if a poleward latitudinal entropy gradient
imposed at the lower boundary (Figure~\ref{sun_ang_vel}c).  Note that the mean
rotation profile and specific entropy variation shown in
Figure~\ref{sun_ang_vel}b,c satisfy Equation~\ref{twbe} in the lower convection
zone.  However, thermal wind balance breaks down in the upper CZ, where $R_o$
can be of order unity locally.  Similar results are obtained by interior
convection simulations
\citep[e.g.,][]{Ellietal00,RobiChan01,BrunToom02,Brunetal04}.  However, the
relative amplitude of the thermal perturbation associated with the
non-cylindrical wind component is $\sim$10$^{-6}$ to $\sim$10$^{-5}$,
corresponding to a very small temperature variation of $\sim$10\,K
\citep{Miesetal06}.  Such variations are too small to be detected by
helioseismology \citep{Gougetal96}.  Hence, at present it is unknown whether
Equation~\ref{twbe} is satisfied in the deep solar CZ.

As was the case for exoplanets, numerical resolution also plays a significant
role.  Low resolution simulations lead to a multi-celled meridional circulation
profiles (not shown); but, with higher resolution, simulations exhibit a
meridional circulation dominated by a single cell in each hemisphere
(Figure~\ref{sun_ang_vel}d).  In the latter, there is a pole-ward flow in the
upper convection zone and an equator-ward flow in the lower CZ.  According to
\citet{Miesetal08}, the thin counter-cells near the lower boundary are likely
the effect of the thermal and mechanical boundary conditions.  Simulations of
solar convection that include an underlying stable zone exhibit equator-ward
circulation throughout the overshoot region as a consequence of the turbulent
alignment of downflow plumes \citep{Miesetal00}.

\adjustfigure{110pt}

\subsection{Some  Important Issues}\label{sun_iss}

Throughout the preceding discussion of the Sun, magnetic fields have essentially
been left out.  This is clearly not justified in the interior.  Significantly,
global MHD convection simulations indicate that dynamo-generated magnetic fields
tend to suppress the rotational shear established by Reynolds stresses and
baroclinicity, which is in contrast to the helioseismology data.  

In some dynamo theories, toroidal magnetic fields are generated and stored in
the tachocline: in these theories, the  meridional flow transports the toroidal
field in the tachocline towards the Equator (which could explain the ``butterfly
diagram'') and the emergence of loops on the surface that form the sunspot
regions.  However, the strength of the return meridional flow is largely unknown
and the required eddy diffusivity is high---roughly 10 times that predicted by
standard mixing-length theory (MLT).  Of course, the accuracy of MLT is also
uncertain.  Another idea for the interior is that proposed by \citet{Balb09},
who argues that weak magnetization renders CZ more prone to baroclinic
instabilities than without magnetization.  Although the problem, in principle,
requires the knowledge of the functional relationship between entropy and
rotation, even simple models readily produce results in broad agreement with the
helioseismology data.  The theory, however, does not apply to the tachocline,
where a simple thermal wind balance may not be valid.

A possible alternative to the above ideas is a model that incorporates the
presence of the near-surface rotational-shear layer, in which the magnetic field
is generated in the bulk of the convection zone but the butterfly pattern is
generated in the shear layer.  Interestingly, local-helioseismology does provide
evidence that the meridional circulation may consist of two-tiered radial
cells. Synoptic analysis of magnetic patterns on the solar surface, such as
rotation of sunspot groups and their inclination relative to the Equator, known
as Joy's law \citep[see, e.g.,][]{ThomWeis08} may provide additional important
information.  Unfortunately, at present the interpretation of
local-helioseismology inversion is a subject of some debate---although steadily
more realistic numerical simulations is alleviating the debate.  

It has been argued above that the solar tachocline plays an essential role in
the solar dynamo, and ultimately the jets, as the likely region in which mean
toroidal flux is generated and stored, eventually emerging from the solar
photosphere as bipolar active regions.  Transport of helical magnetic flux into
the tachocline and the generation of nonhelical fields via rotational shear may
also promote large-scale field generation in the convection zone by
circumventing dynamical quenching constraints.  Furthermore, simulations of
convection indicate that the presence of a tachocline can help organize and
amplify mean fields and can modify the differential rotation profile throughout
the solar envelope via baroclinic forcing.  Poleward angular momentum transport
in the tachocline owing to instabilities or penetrative convection may also
influence the global rotation profile, possibly offsetting equatorward transport
by giant cells in the convection zone \citep{Gilmetal89}.

In summary, the following crucial questions come to the fore: ``How do the upper
and lower boundary layers influence the internal dynamics of the CZ?''  and
``How is the tachocline confined?''  Pertaining to the first question, issues
include baroclinic forcing, gyroscopic pumping, magnetic helicity flux,
tachocline instabilities, and inertial, gravity, and Alfv\'{e}n waves, as well
as the ways in which each affects the thermal, mechanical, and magnetic coupling
between the RC and CZ.  Concerning the second question, the relative roles of
fossil magnetic fields, dynamo-generated fields, tachocline instabilities, and
internal gravity waves need to be better studied.  The tachocline is thought to
be maintained against the downward spreading of differential rotation induced by
gyroscopic pumping, radiative diffusion, and baroclinic
circulations.  Tachocline instabilities \citep{GilmFox97}, fossil magnetic
fields \citep{GougMcIn98}, and internal gravity waves \citep{Taloetal02} may all
act to halt this spreading and thereby to maintain uniform rotation in the
radiative interior.  Understanding this most fundamental question is an
essential prerequisite to understanding the tachocline's broader role in the
dynamics of the solar interior.  Moreover, both have far reaching consequences
for understanding other type of main-sequence stars as well.

\adjustfigure{130pt}

\section{DISCUSSION}\label{discuss}

We conclude this Chapter by summarizing and commenting on some key observations
and simulation results of jets on exoplanets and on and in the Sun.  Experience
from studies of the Earth and other Solar System planets shows that a hierarchy
of theoretical models is necessary to build a robust understanding of both
exoplanets and the Sun, and ultimately to help interpret observations.  Solar
System planets can---and should---be used as validation, as well as a general
guide.  But, we must bear in mind still the many things that are poorly
understood even for the Solar System planets.  For example, what is the deep
structure of the Jovian jets?  Often, the way the atmosphere responds to the
various forcing and damping is a subtle affair, requiring detailed knowledge of
the species composition and distribution; knowledge of the physical properties
of fluids in high pressure environment is critical but still the stuff of
forefront research.  Uranus and Venus starkly show us that the atmosphere can
respond in a manner that is quite unexpected: Uranus has zonal jets, even though
it is heated at the pole, while Venus' cloud top is dominated by an equatorial
jet and a stable polar vortex even though the planet rotates extremely slowly.

On synchronized planets, waves and eddies (not just the mean flow) would likely
be involved in transporting heat between the dayside and the nightside, as well
as between the tropics and the poles.  There is likely to be upwelling motion at
the substellar point and downwelling motion away from this point and inside
polar vortices---if they exist \citep[e.g.,][]{Choetal03}.  Wave momentum flux
and adjustments could reduce shears and temperature gradients, and these
mechanisms remain to be modeled accurately.  As noted, current simulations are
in agreement about the small number of broad jets to be found on synchronized
planets, but they do not resolve small-scale processes nor incorporate realistic
forcing and boundary conditions.  Moreover, ideal and non-ideal MHD simulation
studies are still at the beginning stages.

For the Sun, here we have reviewed many of the subtleties that must be
confronted.  In the past few decades, helioseismology has provided evidence that
CZ is bounded above and below by strong shear layers, and the presence of such
layers can have a critical role in influencing the overall dynamics.  Our
understanding of the solar interior and surface has advanced substantially, due
to helioseismology and continuous observations of solar oscillations by the
Helioseismic and Magnetic Imager (HMI) onboard SDO and by GONG.  High-precision
measurements of oscillation frequencies have provided the radial sound speed
profile and the distribution of the angular velocity through the whole interior,
except perhaps the very inner core of the Sun.  Observations of local processes
steadily becoming available, along with advances in computing, will add to the
remarkable advances that have already been made in this research area.

\vspace*{.5cm}
\noindent{\bf Acknowledgments}\\
We thank Craig Agnor, David Dritschel, Boris Galperin, Mark Miesch, Richard
Scott and Sergei Vorontsov for helpful
discussions.  JWS was supported in part by the grant STFC ST/N504257/1.\\

\bibliography{exo_chap}\label{refs}
\bibliographystyle{cambridgeauthordate}

\end{document}

%% file: chocros.tex
\usepackage{amssymb}
\usepackage{upgreek}
\usepackage{bm}
\usepackage{bbm}
\usepackage{dsfont}
\usepackage{sansmath}
\usepackage{mathrsfs}
\usepackage{yfonts}
\usepackage{euscript}
\usepackage{wasysym}

\def\del{\partial}

\def\grad-s{\mbox{\boldmath{$\nabla$}}_{\!\! s}\,}

\def\bfv{{\bf v}}

\def\bfOmega{\mbox{\boldmath{$\Omega$}}}

\def\gsim{\;\rlap{\lower 2.5pt\hbox{$\sim$}}\raise 1.5pt\hbox{$>$}\;}
\def\lsim{\;\rlap{\lower 2.5pt\hbox{$\sim$}}\raise 1.5pt\hbox{$<$}\;}

\usepackage{color}
\definecolor{com_red}{rgb}{.8,.2,0.1}
\definecolor{ins_green}{rgb}{.1,.5,0.1}